\documentclass[letterpaper,10pt]{article}
\usepackage{authblk}

\usepackage{graphicx}
\usepackage[utf8]{inputenc}
\usepackage[capitalize]{cleveref}
\usepackage{hhline}
\usepackage{comment}
\usepackage{multirow}
\usepackage{enumitem}

\usepackage{pgfplots}
\newif\ifdraft
\newif\ifappnames
\appnamesfalse
\draftfalse

\ifdraft
    \newcommand{\TODOS}[1]{\textbf{Seb:} [ #1 ] }
    \newcommand{\TODOM}[1]{\textbf{Mal:} [ #1 ] }
\else
    \newcommand{\TODOS}[1]{}
    \newcommand{\TODOM}[1]{}
\fi
\definecolor{pincolor}{RGB}{220,87,48}
\definecolor{voicecolor}{RGB}{255,153,0}
\definecolor{notificationcolor}{RGB}{102,140,217}
\definecolor{combicolor}{RGB}{16,150,24}
\definecolor{nearbycolor}{RGB}{175,100,175}
\definecolor{nonecolor}{RGB}{40,178,198}

\definecolor{notatallcolor}{RGB}{175,175,175}
\definecolor{slightlycolor}{RGB}{150,200,200}
\definecolor{moderatelycolor}{RGB}{225,200,100}
\definecolor{verycolor}{RGB}{250,125,125}
\definecolor{extremelycolor}{RGB}{175,100,175}

\makeatletter

\tikzstyle{chart}=[
    legend label/.style={font={\scriptsize},anchor=west,align=left},
    legend box/.style={rectangle, draw, minimum size=5pt},
    axis/.style={black,semithick,->},
    axis label/.style={anchor=east,font={\tiny}},
]

\tikzstyle{bar chart}=[
    chart,
    bar width/.code={
        \pgfmathparse{##1/2}
        \global\let\bar@w\pgfmathresult
    },
    bar/.style={very thick, draw=white},
    bar label/.style={font={\bf\small},anchor=north},
    bar value/.style={font={\footnotesize}},
    bar width=.75,
]

\tikzstyle{pie chart}=[
    chart,
    slice/.style={line cap=round, line join=round, very thick,draw=white},
    pie title/.style={font={\footnotesize\bf}},
    slice type/.style 2 args={
        ##1/.style={fill=##2},
        values of ##1/.style={font={\footnotesize}}
    }
]

\pgfdeclarelayer{background}
\pgfdeclarelayer{foreground}
\pgfsetlayers{background,main,foreground}

\newcommand{\pie}[3][]{
    \begin{scope}[#1]
    \pgfmathsetmacro{\curA}{90}
    \pgfmathsetmacro{\r}{1}
    \def\c{(0,0)}
    \node[pie title] at (90:1.3) {#2};
    \foreach \v/\s in{#3}{
        \pgfmathsetmacro{\deltaA}{\v/100*360}
        \pgfmathsetmacro{\nextA}{\curA - \deltaA}
        \pgfmathsetmacro{\midA}{(\curA+\nextA)/2}

        \path[slice,\s] \c
            -- +(\curA:\r)
            arc (\curA:\nextA:\r)
            -- cycle;
        \pgfmathsetmacro{\d}{max((\deltaA * -(.5/50) + 1) , .5)}

        \begin{pgfonlayer}{foreground}
        \path \c -- node[pos=\d,pie values,values of \s]{$\v\%$} +(\midA:\r);
        \end{pgfonlayer}

        \global\let\curA\nextA
    }
    \end{scope}
}

\newcommand{\legend}[2][]{
    \begin{scope}[#1]
    \path
        \foreach \n/\s in {#2}
            {
                  ++(0,-10pt) node[\s,legend box] {} +(5pt,0) node[legend label] {\n}
            }
    ;
    \end{scope}
}

\newcommand{\makestackedbarchart}[7]{
\begin{tikzpicture}
\begin{axis}[
    ybar stacked,
	bar width=10pt,
    enlargelimits=0.15,
    legend style={at={(0.5,-0.35)},
      anchor=north,legend columns=3},
      legend cell align={left},
    ylabel={},
    symbolic x coords={#1},
    xtick=data,
    x tick label style={rotate=45,anchor=east},
    ]
\addplot+[ybar,color=pincolor,text=black] plot coordinates {#2};
\addplot+[ybar,color=notificationcolor,text=black] plot coordinates {#3};
\addplot+[ybar,color=voicecolor,text=black] plot coordinates {#4};
\addplot+[ybar,color=nearbycolor,text=black] plot coordinates {#5};
\addplot+[ybar,color=combicolor,text=black] plot coordinates {#6};
\addplot+[ybar,color=nonecolor,text=black] plot coordinates {#7};
\legend{\strut pin, \strut notification, \strut voice, \strut nearby devices, \strut combination, \strut none}
\end{axis}
\end{tikzpicture}
}

\newcommand{\makestackedbarchartfromfile}[1]{
\begin{tikzpicture}
\begin{axis}[
    width=15cm,
    height=10cm,
    ybar stacked,
	bar width=4.5pt,
    enlargelimits=0.025,
    legend style={at={(0.5,-0.35)},
      anchor=north,legend columns=3},
      legend cell align={left},
    ylabel={},
    symbolic x coords={Comm1,Comm2,Comm3,Comm4,Comm5,Fin1,Fin2,Fin3,Fin4,Fin5,Prod1,Prod2,Car1,Car2,Car3,Car4,Car5,Dev1,Dev2,Dev3,Dev4,Dev5,Bill1,Bill2,Bill3,Bill4,Bill5,Shop1,Shop2,Shop3,Shop4,Shop5,Health1,Health2,Travel1,Travel2,Travel3,Travel4,Travel5,Music1,Music2,Brief1,Brief2,Edu1,Edu2,Games1,Games2,Food1,Food2,Faith1,Faith2},
    xtick={Comm1,Fin1,Prod1,Car1,Dev1,Bill1,Shop1,Health1,Travel1,Music1,Brief1,Edu1,Games1,Food1,Faith1},
    xticklabels={Communication,Finance,Productivity,Connected Car,Connected Devices,Bills and Utilities,Shopping,Health,Travel,Music,Briefing,Education,Games,Food,Faith},
    x tick label style={rotate=45,anchor=east},
    ]
    \addplot+[ybar,color=pincolor,text=black] table[x=Title,y=pin,col sep=comma] {figures/testplot.txt};
    \addplot+[ybar,color=notificationcolor,text=black] table[x=Title,y=notification,col sep=comma] {figures/testplot.txt};
    \addplot+[ybar,color=voicecolor,text=black] table[x=Title,y=voice,col sep=comma] {figures/testplot.txt};
    \addplot+[ybar,color=nearbycolor,text=black] table[x=Title,y=nearby devices,col sep=comma] {figures/testplot.txt};
    \addplot+[ybar,color=combicolor,text=black] table[x=Title,y=combination,col sep=comma] {figures/testplot.txt};
    \addplot+[ybar,color=nonecolor,text=black] table[x=Title,y=none,col sep=comma] {figures/testplot.txt};
\legend{\strut PIN, \strut Notification, \strut Voice authentication, \strut Nearby devices, \strut A combination of two or more, \strut None}
\end{axis}
\end{tikzpicture}
}
\newcommand{\makedoublepiemeasures}[4]{
\begin{tikzpicture}
[
    pie chart,
    slice type={pin}{pincolor},
    slice type={notification}{notificationcolor},
    slice type={voice}{voicecolor},
    slice type={nearby}{nearbycolor},
    slice type={combination}{combicolor},
    slice type={none}{nonecolor},
    pie values/.style={font={\small}},
    scale=1.35
]
    \pie[xshift=-1cm]{#1}{#2}
    \pie[xshift=2.5cm,values of coltello/.style={pos=1.1}]%
        {#3}{#4}
    \legend[shift={(-2cm,-1cm)}]{{PIN}/pin, {Notification}/notification, {Voice authentication}/voice}
    \legend[shift={(1cm,-1cm)}]{{Nearby devices}/nearby, {A combination of two or more}/combination, {None}/none}
\end{tikzpicture}
}

\newcommand{\makesinglepiemeasures}[2]{
\begin{tikzpicture}
[
    pie chart,
    slice type={pin}{pincolor},
    slice type={notification}{notificationcolor},
    slice type={voice}{voicecolor},
    slice type={nearby}{nearbycolor},
    slice type={combination}{combicolor},
    slice type={none}{nonecolor},
    pie values/.style={font={\small}},
    scale=1.35
]
    \pie[xshift=-1cm]{#1}{#2}
    \legend[shift={(1cm,1.25cm)}]{{PIN}/pin, {Notification}/notification, {Voice authentication}/voice, {Nearby devices}/nearby, {A combination of two or more}/combination, {None}/none}
\end{tikzpicture}
}

\newcommand{\makedoublepieimportance}[4]{
\begin{tikzpicture}
[
    pie chart,
    slice type={notatall}{notatallcolor},
    slice type={slightly}{slightlycolor},
    slice type={moderately}{moderatelycolor},
    slice type={very}{verycolor},
    slice type={extremely}{extremelycolor},
    pie values/.style={font={\small}},
    scale=1.35
]
    \pie[xshift=-1cm]{#1}{#2}
    \pie[xshift=2.5cm,values of coltello/.style={pos=1.1}]%
        {#3}{#4}
    \legend[shift={(-2cm,-1cm)}]{{Not at all important}/notatall, {Slightly important}/slightly, {Moderately important}/moderately}
    \legend[shift={(1cm,-1cm)}]{{Very important}/very, {Extremely important}/extremely}
\end{tikzpicture}
}

\newcommand{\makedoublepiefrequency}[4]{
\begin{tikzpicture}
[
    pie chart,
    slice type={activate}{slightlycolor},
    slice type={newvoice}{moderatelycolor},
    slice type={oncecommand}{verycolor},
    slice type={everycommand}{extremelycolor},
    pie values/.style={font={\small}},
    scale=1.35
]
    \pie[xshift=-1cm]{#1}{#2}
    \pie[xshift=2.5cm,values of coltello/.style={pos=1.1}]%
        {#3}{#4}
    \legend[shift={(-3cm,-1cm)}]{{Only when the user activates it}/activate, {For every new voice}/newvoice}
    \legend[shift={(0.5cm,-1cm)}]{{Once for all commands within one interaction}/oncecommand, {For every command}/everycommand}
\end{tikzpicture}
}

\begin{document}
\date{}
%\title{Security and Usability of Voice Controlled Devices in a Multi-User Setting}
\title{``Tell me, how do you know it's me?''\\Expectations of security and personalization measures for smart speaker applications}
%\title{``Tell me, how do you know it's me?''\\Expectations of access control options for smart speaker applications}

\def\plainauthor{Author name(s) for PDF metadata. Don't forget to anonymize for submission!}

%for single author (just remove % characters)
\author{Maliheh Shirvanian, Sebastian Meiser}
\affil{Visa Research}

%\thecopyright

\maketitle

\begin{abstract}
Voice-controlled smart speaker devices have gained a foothold in many modern households. Their prevalence combined with their intrusion into core private spheres of life has motivated research on security and privacy intrusions, especially those performed by third-party applications used on such devices. In this work, we take a closer look at such third-party applications from a less pessimistic angle: we consider their potential to provide personalized and secure capabilities and investigate measures to authenticate users (``PIN'', ``Voice authentication'', ``Notification'', and presence of ``Nearby devices''). To this end, we asked 100 participants to evaluate 15 application categories and 51 apps with a wide range of functions. The central questions we explored focused on: users' preferences for security and personalization for different categories of apps; the preferred security and personalization measures for different apps; and the preferred frequency of the respective measure.

After an initial pilot study, we focused primarily on 7 categories of apps for which security and personalization are reported to be important; those include the three crucial categories finance, bills, and shopping. We found that ``Voice authentication'', while not currently employed by the apps we studied, is a highly popular measure to achieve security and personalization. Many participants were open to exploring combinations of security measures to increase the protection of highly relevant apps. Here, the combination of ``PIN'' and ``Voice authentication'' was clearly the most desired one. This finding indicates systems that seamlessly combine ``Voice authentication'' with other measures might be a good candidate for future work.

    %Our findings indicate that notifications to phones and voice recognition are widely accepted and that measures taking into account nearby devices have not penetrated public perception enough to be popular.
\end{abstract}

\section{Introduction}

%%define what voice controlled device is and what the ``app/skill'' does

A voice-controlled smart speaker provides a convenient way of monitoring and controlling smart home appliances and running applications for tasks such as online shopping, and playing music and video. Unlike personal smart devices such as smartphones, a smart speaker is typically shared among multiple users, including family members (adults and children), roommates, and occasional visitors such as guests and service personnel. These devices can receive voice commands from anyone over an uncontrolled communication channel. Hence, a natural question is: to which degree can the device ensure it is talking to an authorized user? This question directly ties into the level of personalization the device can provide to legitimate users as well as security and privacy provided in case an unauthorized user interacts with the device. 

We consider this question relevant in a wide variety of cases: roommates or family members might share the same voice-controlled device; friends, or guests might gain physical access to the device; in rare cases, even people without direct physical access to the room of the device can sometimes speak to the device, either remotely via speakers of other devices, or by speaking through a window if they are in proximity of the device. Such concerns are exacerbated with third-party applications running on the device which can have direct access and linkage to payment systems (e.g., to allow users to make purchases or to read out financial account information).

\ifdraft
\begin{itemize}
    \item Alexa Skills!
    \item multi-user setting and security/usability concerns
    \item Motivation (examples)
    \item Research Questions
    \item Contributions
\end{itemize}
\fi

% I think next paragraph (starting from we follow) is not in the right place and only emphasizes one of our questions
\paragraph{This work}
The main goal of this work is to gain an understanding of the users’ perceptions of the security of smart speaker devices and their expectation of task personalization. To this end we ran a user study where we questioned participants familiar with smart speaker devices on their expectations. We followed this up by asking explicitly which security and personalization measures (for access control) the users would prefer for which apps and which frequency of said measures they would feel comfortable with.
%%MSH authentication or access control 

%security and personalization measures
 
%% we can say we selected alexa because in the pre study majority of the users owned this device, also other devices provide similar skills and measures // We should replace task with skill but have a definition of the skill
%% which study
%For our study we considered a smart speaker device. 
Since the questions we targeted rely on non-core functionalities of such devices, we phrased them in the context of third-party applications that can commonly be found and installed on smart speakers. Such apps can add entertainment, allow easy access to information, or connect to other services and devices so they can be voice controlled via the smart speaker.

\paragraph{Contributions}
We ran an extensive user study with 100 participants on security and personalization measures of third-party apps on smart speakers. As part of the study we explored:
\begin{itemize}
    \item Perception of relevance of security and personalization for different categories of apps. 
    Explicitly we asked the participants to rate the level of importance of security and privacy for different application categories. 
    Here, we see that categories around finances and smart devices are most relevant for security, while communication and finance apps are most relevant for personalization. 
    \item Preferred security and personalization measures for different apps. 
    The participants in our user study selected their preferred measure among the four options ``PIN'', ``Notification'', ``Voice authentication'', ``Nearby devices''. 
    Here, for apps traditionally associated with finances or bills, ``PIN'' was the most popular individual measure, while for many other categories ``Voice authentication'' was more popular.
    \item Preferred frequency of the respective measure. Here, we aim to understand how often the user would prefer to (re-)authenticate ranging from ``for every interaction'' to ``only once to activate the app''.
    The most popular frequency choice here was ``once for all the commands within one interaction''.
\end{itemize}
Our study can serve as a baseline for future designs. An especially relevant insight here is that users are comfortable with adding ``Voice authentication'' as a secondary security measure (``Voice'' was included in 4 out of 5 of the most popular combinations of security measures). Given that smart speakers are already controlled via voice, we encourage future designs to include ``Voice authentication'' as a seamless addition to other measures.

\iffalse

To this end, we ran a user study online and asked about users' perception and preferences for the different apps. 
We collect a set of popular apps and for each task, we define the level of customization offered or expected. We then ask users what they expect to be protected or customized on each task. 

Potential concerns we had in mind were, among others:
1)	A malicious entity reserves your room on Airbnb. S/he interacts with the device to access your calendar and gets to know about your upcoming travel. S/he may plan a robbery while you are away.  
2)	Kids may interact with the device and access contents not appropriate for kids or make purchases that are not authorized by an adult. 
3)	You and your roommate share the same voice-controlled device in the kitchen but would like to make online purchases using your card. 
4)	Stretch: you and your spouse have a shared family car and would like to adjust the seat setting for each driver automatically.

Several techniques can be leveraged to protect the security of the app and to allow for higher degree of personalization. Such as ``Voice authentication'', ``PIN'', passphrases, push notification, and pairing with other personal devices.

    While for apps with lower requirements for security and personalization most participants preferred not to use security and personalization measures

\fi

\section{Related Work}

%% MSH should we conclude by saying none have looked at security and personalization as desired by users?

    There is an extensive body of research looking into security and privacy of smart speakers and home assistant devices, broadly falling into 4 categories: works looking into privacy notices and policies \cite{alhadlaq1902privacy,guo2020skillexplorer,edu2021skillvet}, works looking into vulnerabilities and concerns regarding the device itself \cite{lau2018alexa,carlini2016hidden,chung2017digital}, works exploring attack vectors of malicious third-party applications such as \emph{Skills} for Amazon Alexa \cite{su2020you,kumar2018skill,zhang2018understanding,mitev2019alexa,lentzsch2021hey}, and works exploring user perception and preferences \cite{naeini2017privacy,huang2020amazon}. 
    
    \paragraph{Privacy notices}
    Alhadlaq et al.~\cite{alhadlaq1902privacy} take a closer look at the privacy policies of Alexa skills, finding that the vast majority (75\%) lack one entirely and even the ones that present one do not have one customized to smart speaker devices.
    Guo et al.~\cite{guo2020skillexplorer} performed an extensive study on the application behavior of over 30k apps, the vast majority of which from Amazon Alexa. They looked at whether or not the apps access private information besides what was declared in privacy notices and descriptions.
    \cite{edu2021skillvet} automatically analyzed about 200,000 Amazon Alexa skills. Their automated tool, SkillVet, analyzes how well an app's permissions reflect the respective app's published privacy policy.

    \paragraph{General attacks and privacy concerns}
    Lau et al.~\cite{lau2018alexa} examined privacy concerns and perception of smart speakers more generally, without a specific focus on apps. 
    Carlini et al.~\cite{carlini2016hidden} show that hidden voice commands, unintelligible to the human listener, can be used to influence smart speakers.
    Chung et al.~\cite{chung2017digital} examine the possibility of digital forensics in the Amazon Alexa ecosystem; they present a tool for extracting artifacts and studying how investigators can use it to gather information about a users and actions.
        
    \paragraph{Malicious apps}
    Su et al.~\cite{su2020you} explore vulnerabilities of Amazon's Alexa, finding that maliciously created skills/apps can bypass security features to violate user privacy.
    Kumar et al.~\cite{kumar2018skill}, besides studying accidental and systematic misinterpretations of spoken voice commands, propose an attack which they coin \emph{skill squatting}. Here, systematic misinterpretations are leveraged to trick a user into activating a specific skill without their knowledge.
    Zhang et al.~\cite{zhang2018understanding} perform a study on skill squatting attacks on voice assistants, as well as on skill masquerading attacks, where a malicious skill impersonates either the voice assistant or another, legitimate skill with the goal of extracting information.
    Similarly, Mitev et al.~\cite{mitev2019alexa} show how maliciously created skills can hijack the conversation between a user and Alexa; moreover, if a second, malicious IoT device is present, this second device can jam commands, record commands, or perform a man-in-the-middle attack.
    
    Lentzsch et al.~\cite{lentzsch2021hey} perform a large-scale analysis of Amazon Alexa skills and delve into the skill publishing ecosystem. They find and point out methods in which malicious skill developers can publish skills under a fake name (matching existing developers and companies) or significantly modify the backend behavior of a skill after it has been vetted by Amazon.

    \paragraph{User perception and preferences}
    \cite{naeini2017privacy} performed a study on 1,000 participants about their privacy preferences with regards to IoT devices in different scenarios. They find common trends between participants' answers to seemingly different questions pertaining to distinct scenarios; they show how these trends can be used to leverage participants' responses to a few questions to accurately predict how they will respond to other questions.
    
    Huang et al.~\cite{huang2020amazon} present a small-scale study consisting of semi-structured interviews of participants from 21 households. They focused on the experience of shared smart speaker devices.

    %ending it with how we are different

\section{Smart Speaker App Ecosystem}

    \subsection{Apps} 
    Smart speaker apps are offered by app providers and developers to deliver voice-enabled contents to the users. At the time of this study, there were hundreds of thousands of apps offering various experiences ranging from games, music, and entertainment to finance, smart devices, and connected cars. Smart speaker websites categorize the available apps under several main categories depending on the content provided (the complete list of our categories is available in \Cref{sec:classify}). Under each listed app, developers can mention the name, description, security, and permissions required to enable the app and link to the privacy policy document. While this information is helpful to get some understanding of the app, it does not seem to be reviewed by the smart speaker providers and does not necessarily cover all details. Other pieces of information typically available for each app are the number of reviews, a rating, and descriptive reviews written by the users. In \Cref{sec:crawl}, we mention how we used this information to compile a list of apps for our study. 
    
\subsection{Data Collection (App Crawling)}
\label{sec:crawl}
Since most of the apps are offered on all major smart speaker providers we do not focus on any specific smart speaker in this study. However, to collect a representative sample of applications we selected one of the most widely used devices (anonymized for submission) as an example and extracted data from their website without loss of generality. 
We extracted app data from the selected smart speaker website using the WebScraper Chrome browser extension. We defined a sitemap pointing to each of the categories listed on their apps' website. Using WebScraper, for each category we navigated through all pages, crawled each app, and extracted: 1)  App Category 2) App Name, 3) Developer, 4) Description, 5) Number of Reviews, 6) Customer Rating, 7) Account Linking Information, 8) Access and Permissions, and 9) Content Rating (i.e., Mature, Guidance Suggested). We imported the data into a database for further processing.  
%Seb: replaced  Customer's Global Rating by Customer Rating

%As an example, our crawling shows that under the Business and Finance category, a credit card company has a app with around customer reviews, rating the app 3.5 out of 5. This app is rated as Guidance Suggested and enabling it makes the app available on all linked devices. The description provided by developers mentions that using this app users can make payments, check balance and more. As an extra security layer, the app lets users create a 4-digit PIN. 

An example here would be an app named ExampleBanking in the Business and Finance category, created by ExampleDeveloper. The app has 12,000 reviews with an average rating of 3.5 out of 5. The app is rated as Guidance Suggested and enabling it makes the app available on all linked devices. The description provided by ExampleDeveloper mentions that using this app users can make payments, check their balance, and more. As an extra security layer, the app lets users create a 4-digit PIN. 

\subsection{App Categorization and Selection}
\label{sec:classify}

From the total number of 64,000 apps collected using the WebScrapper tool, we selected 230 apps (equal number of apps per each category) with a minimum customer rating of 4.0 and at least 50 reviews. We manually reviewed description and reviews of each of these apps and identified the following: 

\paragraph{App Category} 

%Types of features that the app provides can define the category the app belongs to.
In our study, we wanted to present the participants with a variety of apps that offer distinct and relevant functionalities. An intuitive approach to select such apps would be to rely on app store provider categories and select a few apps from each of those directly. This approach, however, faces challenges. First, those categories are generally not disjoint; the same app may well appear in two or more categories. Second, the categories do not directly distinguish apps by their functionality and rather describe the expected context in which an app might be used. For example, a ``finance and business'' category includes briefing apps and appointment schedulers in addition to banking apps, shopping help, and investment apps.

To better understand how security and personalization features interact with the functionality provided by apps, we decided to create our own categorization of apps that is subtly different from that of app stores. We chose to use the following set of 15 skill categories:
Finance, Utilities/Bill, Communication, Briefing, Faith, Smart Devices, Connected Cars, Shopping, Travel, Productivity, Health, Games, Food, Educational, and Music.

We then assigned each app exactly one of those categories (to keep them disjoint) based on the app's functionality. For example, a rideshare app we considered was listed under ``Local'' as well as ``Travel and transportation''. Several of the food takeout and delivery apps such as a pizza order app are also listed under ``Local''. Given the description of the app, we categorize ride share apps as ``Travel'' while assigning the pizza order app to the ``Food'' category.

%To categorize the skills, we first manually examined 10 of the most highly ranked and most downloaded skills from each of Amazon's 23 categories. For each skill, we determined the skill category manually.

Following is the list of app categories and descriptions of each as included in our user study survey: 

%\label{sec:list-of-categories}
\begin{itemize}
    \item Finance: access your financial information, receive and transfer funds, get balance, access your transactions, stock prices, budgeting.
    \item Utilities/Bill: pay bills, balance info, schedule payment, outage info
    \item Communication: access your contacts, call and send messages, read notifications, emails.
    \item Briefing: listen to news and briefings, weather forecast
    \item Faith: Quotes of the day, listen to inspirational speech, scripture, verses 
    \item Smart devices: access your connected devices, set temperature, lock/unlock door, turn on/off camera, read sensors
    \item Connected Cars: start/stop car engine, lock/unlock car door, get odometer/fuel/temp/location information
    \item Shopping: order items, track orders, sales, and coupons
    \item Travel: get schedules, order ride, access ride history, get direction, route info 
    \item Productivity: calendar, lists, reminders, count downs
    \item Health: workout routines, meditations, calming songs, sleep aid, general health and medical knowledge
    \item Games: trivia, games, fun facts, quizzes, jokes and humour, sport teams fans
    \item Food: food recipes, cocktail recipes, cooking temperature, cooking time, diet, nutrition info
    \item Educational: facts, learning a skill, getting advice, personal growth, common knowledge
    \item Music: playing musics, podcasts and audio-books, relaxing sounds
\end{itemize}

\paragraph{Descriptive Keywords}

For each of our categories, we noted down the most common keywords occurring in the description of the respective apps. We used these keywords as labels to build our automated app categorization to classify the remaining apps by the frequency of occurrence of those keywords as discussed in \Cref{sec:classify}. As an example, keywords such as ``learn'', ``skill'', ``knowledge'', and ``develop'', used in the description of the app suggests that the app offers educational content.  

%Finally, we automatically classified the remaining apps by the frequency of occurrence of those keywords.

\paragraph{Security and Personalization} 

We intend to understand the types of security, personalization, and parental control features a app offers and whether it meets users' perception and expectations. Some of the apps explicitly mention presence or lack of these features in their description (e.g., none, PIN for all commands, PIN for certain commands, voice recognition). Consumer reviews were another useful resource to understand whether such features are available and functional. For apps that security and personalizing seem to be critical, e.g., banks and financial institutions, we enabled the skill on the respective smart speaker device and verified the security or personalization features.

\section{Study Methodology and Design}
    \subsection{Study Setup}
    We ran a small initial pilot study (with 25 participants), followed by a more extensive study with 100 participants. Both studies were hosted on the Cvent Web Survey platform and participants were recruited through Amazon Turk. 
    We required the participants to be Mechanical Turk Masters and requested the participants to take the survey only if they are fully familiar with smart speaker devices.
    The pilot study took on average 23 minutes to complete and the participants were awarded \$2 for their effort. The main study took on average 42 minutes and the participants were compensated \$3.5 for their effort. All participants were given detailed information about the goal of the study and were educated about the terms used in the study, e.g., smart speaker, app, app category, security, personalization, etc. 
    
    \subsection{Ethical Considerations}
    This study attempted to get a sense of people's preferences and intuitions when it comes to security and security measures, as well as personalization and personalization measures of smart speaker devices. We informed all participants about the types of questions we would ask and they were given the option to withdraw from the study at any time. We followed user study best practices to make sure all information collected are kept strictly confidential, only referenced with a non‐identifying code, and reported as aggregate. 
    Our aim is not to individually target a specific smart speaker or app; hence, to protect the privacy of smart speaker providers and app developers, we anonymized the brand names in reporting the aggregated results in this manuscript. The participants, however, were given the name of a smart speaker and names for the apps to provide informed responses. 
    The study was reviewed and approved by our Risk Management and Privacy Center. 
    
    \subsection{Study Questions}
    We arranged the questions into four groups and presented them to each participant in the following order. To avoid bias the ordering of questions within a group was randomized.  
    
    \paragraph{\underline{Group 1.} Background and demographics.}     We started the study with an initial set of questions to determine demographic information as well as prior experience with voice assistants and smart speakers. We also asked about the participant's household composition, the primary owner, other non-primary consumers, and occasional users of the smart home device. 
    
    \paragraph{Questions.}
    \begin{enumerate}[noitemsep]
        \item  What is your gender (Female; Male; Other)?
        \item  Which category below includes your age (18-24; 25-34; 35-44; 45-54; 55-64; 65+)?
        \item  What is the highest level of education you have completed (High school degree or equivalent (e.g., GED); Some college but no degree; Associate degree; Bachelor degree; Graduate degree)?
        \item  Is your major or primary job in information technology, computer science, or computer engineering? If yes, please specify your field (Yes; No; Other).
        \item  How do you describe your "General Computer" skills (Very poor; Poor; Acceptable; Good; Very Good)?
        \item  How do you describe your "General Computer Security" skills  (Very poor; Poor; Acceptable; Good; Very Good)?
        \item  How familiar are you with Home Assistant devices such as a Amazon Alexa and Google Assistant (Not at All Familiar; Slightly Familiar; Moderately Familiar; Very Familiar; Extremely Familiar)?
        \item  Do you own a home assistant device (Yes, Amazon Alexa; Yes, Google Assistant; Yes, Others; No)?
        \item  Who is in your household (Just me; Roommates; Kids; Partner; Parent; Occasional Guests)?
        \item  Is your device shared among multiple family members (Yes, and I am the primary user; Yes, but I am not the primary user; No, I am the only user;)?
    \end{enumerate}

    \paragraph{\underline{Group 2.} Perception and expectation of security and personalization offered by each category.}  Subsequently, we presented the following app categories, asking participants to assess whether they feel ``Secure Access'', ``Personalization'', or ``Parental Control'' is desired for each category. Then we asked how important they would rate a) security and b) personalization for each app category. These questions reveal the overall priorities of each study participant.
    
    \paragraph{Questions.}
    
    \begin{enumerate}[noitemsep]
         \item Please review the following app categories and decide if you would like to see any security or personalization to be offered. Checkboxes: Secure access, Personalization, Parental Control, No security or personalization
        \item For each of the following categories, rate the level of security needed. (Not at all Important;	Slightly Important;	Moderately Important;	Very Important;	Extremely Important)
        \item For each of the following categories, rate the level of personalization needed. (Not at all Important;	Slightly Important;	Moderately Important;	Very Important;	Extremely Important)
    \end{enumerate}
    
    \paragraph{\underline{Group 3.} Desired security and personalization measure  for each presented app.}  The bulk of the questions then consisted of the participants choosing their favorite security and personalization measures and the frequency the app should require the measure to re-authenticatethe user. 
    
    Based on the responses we received in our pilot study, we considered the 7 categories \footnote{Highly relevant Categories: Finance, Communication, Connected Cars, Utilities/Bill, Smart Devices, Shopping, and Travel} that were rated 3 and above in both security and personalization as highly relevant categories. We considered the 8  remaining categories \footnote{Less relevant Categories: Productivity, Health, Music, Educational, Briefing, Games, Food, and Faith} that received a ranking of below 3 in both security and personalization, as less relevant categories.   
    
    We selected a total of 51 apps, 5 apps belonging to each of the 7 highly relevant categories and 2 apps for each of the 8 less relevant categories. We presented the apps' name and description and asked the participants to select their preferred method/s personalization and security measures. While most smart speakers' main measure is PIN, we considered 3 other possible measures as well and described them as follows: 

    \begin{itemize}

    \item PIN: memorize and speak a secret PIN to access those apps that require security or personalization. Anyone knowing the PIN can access the apps. 

    \item Voice authentication: The smart speaker can recognize authorized voices and secure or customize based on the speaker's voice. Replayed voices or a similar voice may be able to gain access. 

    \item Nearby devices: The smart speaker can identify devices such as personal smart phones and smart watches or even a secure key fob and authorize access to the apps only if certain devices are available. If your devices are nearby the apps can be activated by anyone at home. 

    \item Notification: The smart speaker can send a notification message to your smart phone and ask for permission to run sensitive apps. You need to have your phone handy when running such apps.
  
    \end{itemize}
    
    \paragraph{Questions.}    
    \begin{enumerate}[noitemsep]
        \item Which one of the following access control methods you would prefer to use if it is offered by the apps (a list of a few example apps):  ``PIN'', ``Voice authentication'', ``Nearby devices'', ``Notification'', ``None'', ``A combination of two or more''
        \item In case for some of the apps you selected "a combination", what combination of techniques would you prefer to see? Please select two or more.
        \item How frequently do you think the app should require the access control method (e.g., ``PIN'', ``Voice authentication'') if security is the main objective. (For every command; Once for all commands within one interaction; For every new voice; Only when the user activates it; Other )
        \item How frequently do you think the app should require the access control method (e.g., ``PIN'', ``Voice authentication'') if usability is the main objective. (For every command; Once for all commands within one interaction; For every new voice; Only when the user activates it; Other )
    \end{enumerate}
    
    \paragraph{\underline{Group 4.} Open-ended and feedback questions. }
    We ended the study with three open-ended questions, asking the participant to name and describe a app that might benefit from security or personalization feature and provide any additional feedback.  

    \paragraph{Questions.}
    \begin{enumerate}[noitemsep]
        \item Considering the security as the main objective of an app imagine at least one app that can benefit from access control feature. Name or describe the app.
        \item Considering the personalization as your main objective, imagine at least one app that can benefit from access control feature. Name or describe the app.
        \item If any of your answers require additional explanation or you would like to give suggestion and feedback about any part of the study please enter it here.
    \end{enumerate}

    \subsection{Ensuring Validity of Responses}
    To ensure that each participant paid due diligence we added one or more dummy questions to each block of questions and explicitly asked the participant to select a given response. We also manually reviewed the open-ended questions posed to the participants at the end of the study; we discarded those responses that indicated participant negligence. For example, in answer to Question 1 in Group 4 (asking the participant to name or describe an app), we discarded a participant that responded ``Nice Survey''. After reviewing the responses from all 146 participants we discarded all that failed to respond correctly to the dummy questions or to the open-ended questions, leaving us with exactly 100 participants.~\footnote{The turnover rate of the survey was rather short. After discarding some of the initial participants as described above, we reopened the survey once more so that we would have at least 100 valid responses overall. After discarding invalid responses we coincidentally ended up with precisely 100 responses, which made calculating percentages pleasantly simple.}

    \subsection{Pilot Study}
    Our pilot study was a basis to examine the design of our final study and estimate the average time it takes for the participants to respond. In our initial pilot study, we presented the app categories to our participants and asked them to evaluate their relevance for security and personalization. 
    
    As the participants of the pilot study needed around 20 minutes to finish their response (for a reduced set of questions including just 10 specific apps) we decided to limit the number of apps in the final study to avoid response fatigue. To this end, we split our categories into 7 very relevant categories (Finance, Utilities/Bill, Communication, Smart Devices, Connected Cars, Shopping, and Travel) and 8 less relevant categories based on the importance of security or personalization reported by the pilot study participants. While our pilot study only included 10 apps we extended the number of apps to 51 in the final study, five from each of the seven very relevant categories and two from each of the less relevant categories. Moreover, we added dummy questions, open-ended questions, and question about the household composition and ownership of smart home assistant speakers in the final study. 
    
    %In our final study, we included five representative and very popular apps from each of the seven very relevant categories (Finance, Utilities/Bill, Communication, Smart Devices, Connected Cars, Shopping, and Travel), as well as two apps from each of the less relevant categories, for a total of 51 apps.

    \section{Study Results and Analysis}
    \label{sec:results}
    %\paragraph{Responses}
    %\TODOS{Describe how many responses we had to exclude and why}

    \subsection{Demographics and Background - Group 1}
    %\paragraph{Mechanical Turk}

    %\paragraph{Demographics}

\iffalse

\begin{figure}[h]
\centering
\includegraphics[width=0.225\textwidth]{figures/demographics-computer-science-skills}
\includegraphics[width=0.225\textwidth]{figures/demographics-computer-security-skills}
\caption{Demographics: self reported skills in computer science and computer security}\label{fig:demographic-skills}
\end{figure}

\begin{figure}[h]
\centering
\includegraphics[width=0.225\textwidth]{figures/background-which-device.png} \\
\includegraphics[width=0.225\textwidth]{figures/background-do-you-share.png}
\includegraphics[width=0.225\textwidth]{figures/background-household.png}
\caption{Background: Information about the participant's household and smart device.}\label{fig:demographics-devices}
\end{figure}

\fi

\begin{table}[t]
\caption{Demographics Information}
\label{table:demographics}
\scriptsize
\center
\begin{tabular}{|lll|l|lll|}
\cline{1-3} \cline{5-7}
\multicolumn{3}{|l|}{\textbf{Gender}}    &  & \multicolumn{3}{l|}{\textbf{Employment}}                  \\ \cline{1-3} \cline{5-7} 
 & Female                     & 38 &  &  & CS         & 31 \\
 & Male                       & 61 &  &  & Non-CS     & 69 \\
 & Other                      & 1  &  &  & Others     & 0  \\ \cline{1-3} \cline{5-7} 
\multicolumn{3}{|l|}{\textbf{Age}}       &  & \multicolumn{3}{l|}{\textbf{General CS Background}}       \\ \cline{1-3} \cline{5-7} 
 & 18-24                      & 2  &  &  & Very poor  & 0  \\
 & 25-34                      & 46 &  &  & Poor       & 9  \\
 & 35-44                      & 29 &  &  & Acceptable & 43 \\
 & 45-54                      & 13 &  &  & Good       & 48 \\
 & 55-64                      & 9  &  &  & Very Good  & 0  \\
 & 65+                        & 1  &  &  &            &    \\ \cline{1-3} \cline{5-7} 
\multicolumn{3}{|l|}{\textbf{Education}} &  & \multicolumn{3}{l|}{\textbf{General Security Background}} \\ \cline{1-3} \cline{5-7} 
 & High school or equivalent  & 15 &  &  & Very poor  & 3  \\
 & Some college but no degree & 10 &  &  & Poor       & 28 \\
 & Associate degree           & 9  &  &  & Acceptable & 50 \\
 & Bachelor degree            & 59 &  &  & Good       & 19 \\
 & Graduate degree            & 7  &  &  & Very Good  & 0  \\ \cline{1-3} \cline{5-7} 
\end{tabular}
\end{table}

\begin{table}[t]
\caption{The participants household and smart devices.}
\label{table:demographics-devices}
\scriptsize
\center
\begin{tabular}{|lll|l|lll|}
\cline{1-3} \cline{5-7}
\multicolumn{3}{|l|}{\textbf{Familiarity with Smart Speaker}} &  & \multicolumn{3}{l|}{\textbf{Household}}        \\ \cline{1-3} \cline{5-7} 
 & Not at All Familiar & 0  &  &  & Just me              & 20 \\
 & Slightly Familiar   & 1  &  &  & Roommates            & 9  \\
 & Moderately Familiar & 9  &  &  & Kids                 & 8  \\
 & Very Familiar       & 43 &  &  & Partner              & 52 \\
 & Extremely Familiar  & 47 &  &  & Parent               & 6  \\ \cline{1-3}
\multicolumn{3}{|l|}{\textbf{Device Used}}            &  &  & Occasional Guests & 1 \\ \cline{1-3} \cline{5-7} 
            & Amazon Alexa            & 76            &  & \multicolumn{3}{l|}{\textbf{Shared Device}}    \\ \cline{5-7} 
 & Google Assistant.   & 20 &  &  & Yes/Primary User     & 69 \\
 & Others              & 1  &  &  & Yes/Non-Primary User & 10 \\
 & None                & 3  &  &  & Not shared           & 21 \\ \cline{1-3} \cline{5-7} 
\end{tabular}
\end{table}

    Among the valid 100 respondents, 38 identified as female, 61 as male, and 1 as other. The largest age group in our study was people 25-34 years of age, with subsequent groups shrinking in size in clear relation with the age (2\% 18-24 , 46\% 25-34, 29\% 35-44, 13\% 45-54, 9\% 55-64 and 1\% over 65). The majority of participants reported having a bachelor's degree (15\% High school, 10\% Some College, 9\% Associate Degree, 59\% Bachelor's Degree, and 7\% Graduate Degree). 
    31\% of the participants had a computer science background, and majority of the participants declared to have above average computer science and computer security background.   %We refer to \Cref{fig:demographic-age,fig:demographic-education,fig:demographic-skills} for a visual representation of our demographics.
    We refer to \Cref{table:demographics} for a breakdown of our demographics. %and to \Cref{fig:demographic-skills,fig:demographics-devices} for a visual representation of the computer skill background and device usage.

    We asked the participants to only respond to the survey if they felt fully familiar with smart home assistant speakers. Consequently, the majority of participants  had above average familiarity, with 47\%  extremely familiar, 43\% very familiar, 9\% moderately familiar, and 1\$ slightly familiar with such a device. 
    As far as the household composition and ownership of a smart speaker device, the vast majority of the participants responded that they personally owned such a device, with just 3\% of participants responding that they do not. 76\% of participants reported owning an Amazon Alexa, 20\% owning Google Home, and 1\% a non-listed device. 69\% of the participants were the primary user of a shared device, 10\% were non-primary user, and 21\% were the only user of the device. About half of the participants (53\%) responded that they lived with their partner, 20\% lived alone, 9\%  had roommates, 9\% had kids at home, 7\% lived with parents, and 1\% lived with occasional guests such as temporary residents.  
    
    Responses to the demographic questions show that our participants are among the young and educated population who have high understanding of the technology. 
    
  %      \subsubsection{Location, Household Members, Usage Context}
        \subsection{Importance of Security and Personalization -- Group 2, Question 2 and 3}
        \label{sec:importance}

        \begin{table}[]
    \centering
\caption{Average rating of the importance of security and personalization for each category; possible choices ranged from 1 (``not at all important'') to $5$ (``extremely important'').}
\label{table:rating-security-personalization}
        \small{
\begin{tabular}{|ll|l|ll|}
\cline{1-2} \cline{4-5}
\multicolumn{2}{|c|}{\textbf{Importance of Security}} &  & \multicolumn{2}{c|}{\textbf{Importance of Personalization}} \\ \hhline{==~==} %\cline{1-2}\cline{4-5} 
\multicolumn{1}{|l|}{\textbf{Finance}}           & 4.69    &  & \multicolumn{1}{l|}{\textbf{Communication}}         & 4.11       \\ \cline{1-2} \cline{4-5} 
\multicolumn{1}{|l|}{\textbf{Utilities/Bill}}    & 4.10    &  & \multicolumn{1}{l|}{\textbf{Finance}}               & 3.97       \\ \cline{1-2} \cline{4-5} 
\multicolumn{1}{|l|}{\textbf{Connected Cars}}    & 4.08    &  & \multicolumn{1}{l|}{\textbf{Productivity}}          & 3.67       \\ \cline{1-2} \cline{4-5} 
\multicolumn{1}{|l|}{\textbf{Smart Devices}}     & 3.85    &  & \multicolumn{1}{l|}{\textbf{Connected Cars}}        & 3.65       \\ \cline{1-2} \cline{4-5} 
\multicolumn{1}{|l|}{\textbf{Communication}}     & 3.84    &  & \multicolumn{1}{l|}{\textbf{Smart Devices}}         & 3.62       \\ \cline{1-2} \cline{4-5} 
\multicolumn{1}{|l|}{\textbf{Shopping}}          & 3.49    &  & \multicolumn{1}{l|}{\textbf{Utilities/Bill}}        & 3.55       \\ \cline{1-2} \cline{4-5} 
\multicolumn{1}{|l|}{\textbf{Travel}}            & 2.86    &  & \multicolumn{1}{l|}{\textbf{Shopping}}              & 3.51       \\ \cline{1-2} \cline{4-5} 
\multicolumn{1}{|l|}{\textbf{Productivity}}      & 2.60    &  & \multicolumn{1}{l|}{\textbf{Health}}                & 3.24       \\ \cline{1-2} \cline{4-5} 
\multicolumn{1}{|l|}{\textbf{Health}}            & 2.47    &  & \multicolumn{1}{l|}{\textbf{Travel}}                & 3.19       \\ \cline{1-2} \cline{4-5} 
\multicolumn{1}{|l|}{\textbf{Music}}             & 2.05    &  & \multicolumn{1}{l|}{\textbf{Music}}                 & 3.14       \\ \cline{1-2} \cline{4-5} 
\multicolumn{1}{|l|}{\textbf{Educational}}       & 1.97    &  & \multicolumn{1}{l|}{\textbf{Briefing}}              & 2.66       \\ \cline{1-2} \cline{4-5} 
\multicolumn{1}{|l|}{\textbf{Games}}             & 1.95    &  & \multicolumn{1}{l|}{\textbf{Educational}}           & 2.53       \\ \cline{1-2} \cline{4-5} 
\multicolumn{1}{|l|}{\textbf{Briefing}}          & 1.81    &  & \multicolumn{1}{l|}{\textbf{Games}}                 & 2.49       \\ \cline{1-2} \cline{4-5} 
\multicolumn{1}{|l|}{\textbf{Food}}              & 1.81    &  & \multicolumn{1}{l|}{\textbf{Food}}                  & 2.45       \\ \cline{1-2} \cline{4-5} 
\multicolumn{1}{|l|}{\textbf{Faith}}             & 1.69    &  & \multicolumn{1}{l|}{\textbf{Faith}}                 & 2.31       \\ \cline{1-2} \cline{4-5} 
\end{tabular}
}
\end{table}

\begin{figure}[h]
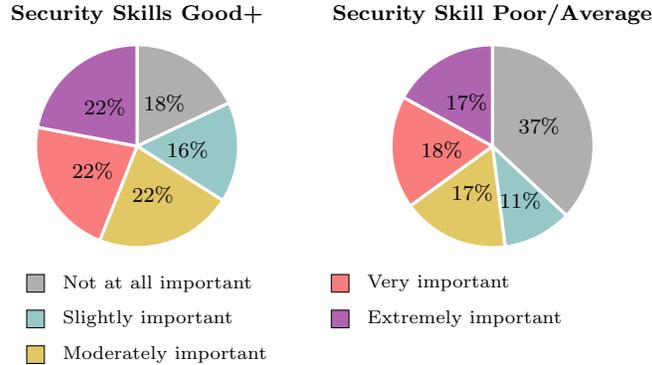

\centering
\makedoublepieimportance{Security Skills Good+}{18/notatall,16/slightly,22/moderately,22/very,22/extremely}{Security Skill Poor/Average}{37/notatall,11/slightly,17/moderately,18/very,17/extremely}
\caption{How important is security, comparison between participants self reporting to have good or very good computer security skill vs participants self reporting to have average or poor computer security skill.}\label{fig:security-comparing-good-security-skill-vs-poor}
\end{figure}

\begin{table}[]
\small
\centering
\caption{Average importance of security and personalization by household configuration.}
\label{table:importance-security-personalization-household}
\begin{tabular}{|l|l|l|}
\hline
\textbf{Household} &
  \multicolumn{1}{c|}{\textbf{\begin{tabular}[c]{@{}c@{}}Importance of \\ Security\end{tabular}}} &
  \textbf{\begin{tabular}[c]{@{}l@{}}Importance of \\ Personalization\end{tabular}} \\ \hline
\textbf{Just me}           & 2.94 & 3.13 \\ \hline
\textbf{My Spouse/Partner} & 3.00 & 3.26 \\ \hline
\textbf{My Roommate(s)}     & 2.80 & 2.73 \\ \hline
\textbf{My Parents}        & 2.48 & 3.05 \\ \hline
\textbf{Kids}              & 2.74 & 3.76 \\ \hline
\textbf{Occasional Guests} & 3.00 & 3.27 \\ \hline
\end{tabular}
\end{table}
        
        To recall, we presented 15 app categories to the participants and asked them to rate the importance of security and personalization for each of the listed categories on a scale of 1 to 5 with 1 representing ``Not at all important'' and 5 representing ``Extremely important''. 
        They rated apps related to \emph{Educational content}, \emph{Games}, \emph{Briefing}, \emph{Food}, and \emph{Faith} to have lower importance with respect to both security and personalization. \Cref{table:rating-security-personalization} summarizes the responses to these two questions. Intuitively app categories with high security and personalization importance would benefit from security and personalization measures to fulfill users' expectations. As an example, it would be desired for an app that reads out bank statements to ask for user authentication before providing the information, while a briefing app that announces the latest news would not require such authentication.
        
        Overall we see that the participants considered security important for a small selection of categories, with the importance falling steeply for other categories (average rating over all categories $2.88$ with a variance of $1.06$). By contrast, participants overall considered personalization to be important, with less nuance between the categories (average rating over all categories $3.21$ with a variance of $0.35$). This finding is explained by the general benefit of personalization for a wide variety of apps, even if those might not be security-critical in nature.
        
        People that self reported to have ``Good'' computer security skills (rating >= 4), rated security to be of more importance compared to those participants who self reported to have ``Acceptable'', ``Poor'', or ``Very Poor'' computer security skills (rating <=3). This result is captured in \Cref{fig:security-comparing-good-security-skill-vs-poor}.

        People who self-reported good computer security skills generally avoided stating that security is ``Not at all important'', instead distributing their assessment over the different categories evenly. A possible explanation for these results is that having self-reported good computer security skills might have primed participants to view the categories with a more critical eye.
    
        We initially had the hypothesis that household compositions might have a significant impact on how important participants rate security and personalization; the idea being that people living in households with children or occasional guests might require very different degrees of security and personalization than people living alone or with a partner only. However, we did not find a significant difference in their responses to the importance of security or personalization as shown in \Cref{table:importance-security-personalization-household}.   
        
        %For people living alone vs not alone: How important is security and personalization? 

\begin{figure}[h!]
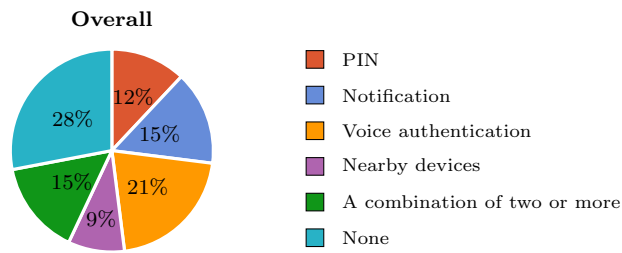

\centering
\makesinglepiemeasures{Overall}{12/pin,15/notification,21/voice,9/nearby,15/combination,28/none}
\caption{Preferred security and personalization measures overall.}\label{fig:measures-overall}
\end{figure}

\begin{figure}[h]
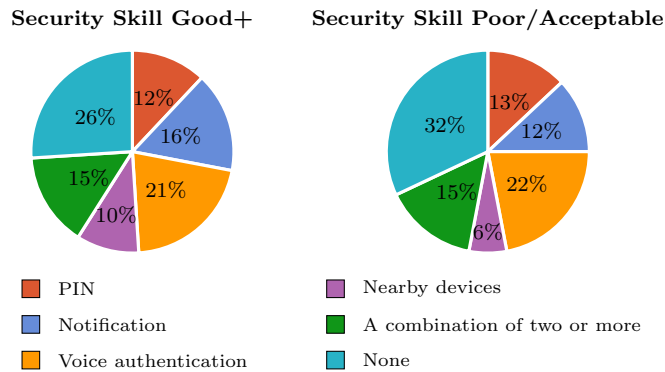

\centering
\makedoublepiemeasures{Security Skill Good+}{12/pin,16/notification,21/voice,10/nearby,15/combination,26/none}{Security Skill Poor/Acceptable}{13/pin,12/notification,22/voice,6/nearby,15/combination,32/none}
\caption{Preferred security and personalization measures overall, comparison between participants self reporting to have good or very good computer security skill vs participants self reporting to have average or poor computer security skill.}\label{fig:measures-comparison-good-security-skill-vs-poor}
\end{figure}

\begin{figure}[h]
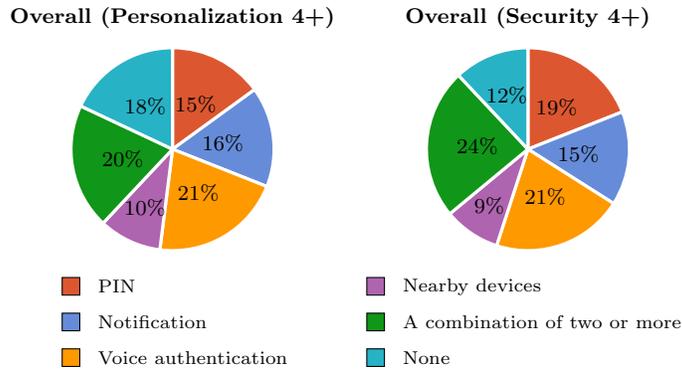

\centering
\makedoublepiemeasures{Overall (Personalization 4+)}{15/pin,16/notification,21/voice,10/nearby,20/combination,18/none}{Overall (Security 4+)}{19/pin,15/notification,21/voice,9/nearby,24/combination,12/none}
\caption{Preferred security and personalization measures selected for apps in categories that the participant rated as 4+ out of 5 in terms of importance for personalization and security respectively.}
\label{fig:measures-skills-security-important-vs-personalization-important}
\end{figure}

\subsection{Security and Personalization Measures - Group 3, Question 1}

\begin{figure*}[h!]
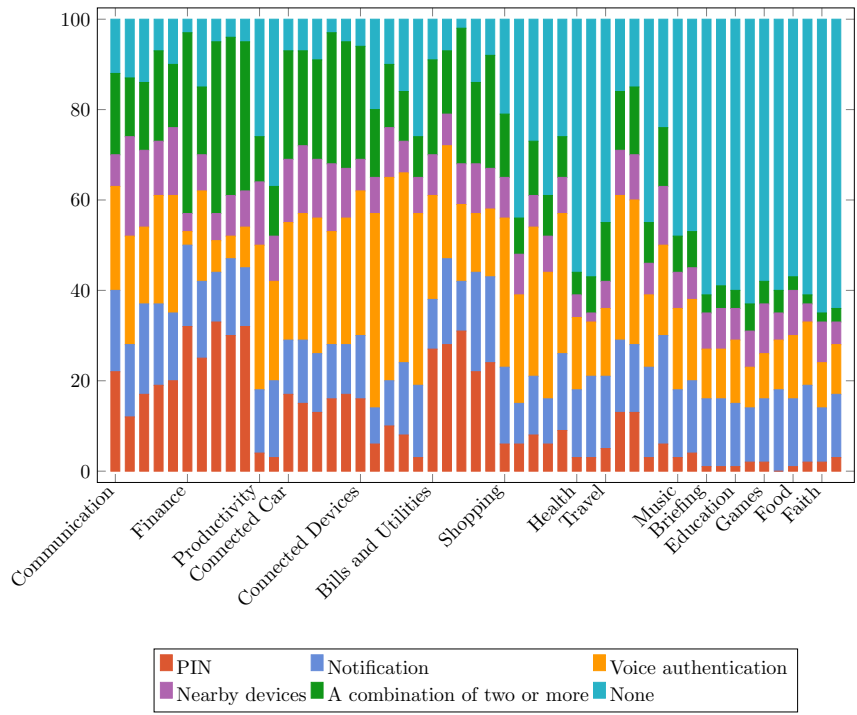

\centering
\scalebox{.75}{
\makestackedbarchartfromfile{figures/TFig5.csv}
}
\caption{Preferred security measure per app over all participants, the categories are sorted by the combined weight of security and personalization.}\label{fig:options}
\end{figure*}

This group of questions asked for the participants' favorite security and personalization measure for each listed app in highly relevant and less relevant categories.   
        
The responses to this set of questions showed that while the participants had heterogeneous preferences, when applied individually, ``Voice authentication'' was rated as the most desirable option overall followed by ``Notification''. 15\% of the responses were in favor of ``A combination of two or more''.  28\% of the responses were in favor of ``None'' (no security or personalization needed). For these percentages we took the average over all apps in our study, including both the highly relevant and the less relevant app categories. Overall, fewer participants were in favor of ``Nearby devices'' and ``PIN''. As mentioned earlier, smart speaker apps only offer PIN when a security or personalization measure is offered at all, which is in contrast to our findings about users' preferences. \Cref{fig:measures-overall} summarizes the  results. \Cref{fig:measures-comparison-good-security-skill-vs-poor} segments the result based on the participants' self reported security skills; we did not notice a large difference in the overall preference of measures for these two groups. Participants self-reporting to have good security skills selected ``None'' less frequently; they also seemed more open to the idea of ``Notification'' and ``Nearby devices'' than their peers. This perhaps indicates that those measures are not yet widely known and could present an opportunity for informing.

    The preferences change slightly when we take context into account: For apps in categories that the participants themselves considered security critical (Group 2 -- Question 2), ``A combination of two or more'', ``Voice authentication'', and ``PIN'' were most often selected. For apps in categories where the participants themselves rated personalization to be important (Group 2 -- Question 3), the measure most often selected was ``Voice authentication''; here ``PIN'' was comparatively unpopular (c.f., \Cref{fig:measures-skills-security-important-vs-personalization-important}).
    
    In \Cref{fig:options} we show the selected measure chosen by participants for each of the individual apps, sorted by the importance of security and personalization of the respective category. Here, we see that even though participants considered security to be similarly important for apps in Finance and Bills as for apps in Smart devices and Connected Car, the preferred measures are drastically different. One possible explanation might be that the participants associated ``PIN'' as a security measure with financial and banking scenarios and thus often selected it as a measure. For connected devices (cars or otherwise), ``Voice authentication'' was by far the preferred measure.
    As expected the participants mostly selected None (i.e., no measure needed) for apps in categories less relevant to security and personalization.

\ifdraft       
    For the top 5 apps where security and personalization is considered very important or extremely important: which method of protection did people choose? + 
    I did the top 5 and determined which those were by ranking them via “number of people selecting “None”, descending. The top 5 were apps with only up to 3 “None” responses.  
\fi

      \subsection{Security and Personalization Measure Combinations - Group 3, Question 2}
      
      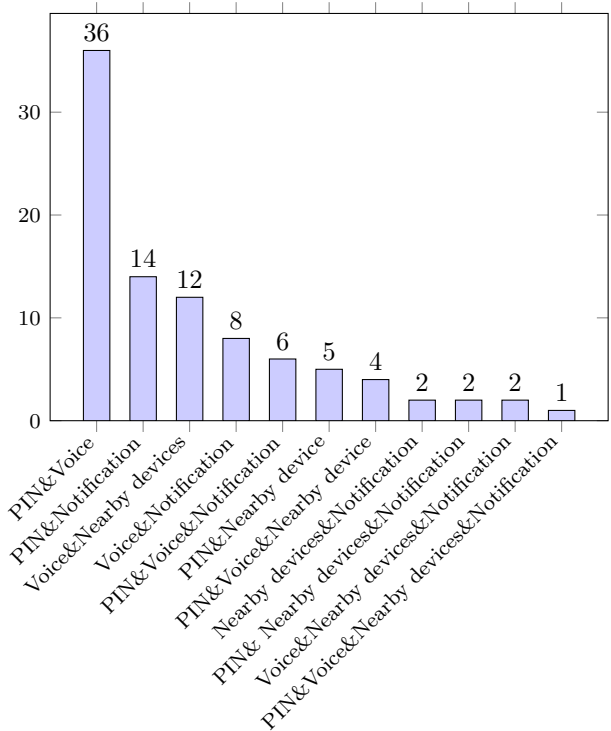
\begin{figure}[h!]
        \centering
        \begin{tikzpicture}
        \begin{axis}[
            width=9cm,
            height=7cm,
            ybar,
        	bar width=10pt,
        	nodes near coords,
            legend style={at={(0.5,-0.35)},
              anchor=north,legend columns=3},
              legend cell align={left},
            ylabel={},
            ymin=0,
            ticklabel style = {font=\footnotesize},
            symbolic x coords={pv,pn,vnd,vn,pvn,pnd,pvnd,ndn,pndn,vndn,pvndn},
            xtick=data,
            xticklabels={PIN\&Voice, PIN\&Notification,Voice\&Nearby devices,Voice\&Notification, PIN\&Voice\&Notification, PIN\&Nearby device, PIN\&Voice\&Nearby device, Nearby devices\&Notification, PIN\& Nearby devices\&Notification, Voice\&Nearby devices\&Notification, PIN\&Voice\&Nearby devices\&Notification},
            x tick label style={rotate=45,anchor=east}
            ]
        \addplot [ybar,fill=blue!20!white,text=black] plot coordinates {(pv,36)(pn,14)(vnd,12)(vn,8)(pvn,6)(pnd,5)(pvnd,4)(ndn,2)(pndn,2)(vndn,2)(pvndn,1)}; %color=pincolor
        %\legend{\strut pin, \strut notification, \strut voice, \strut nearby devices, \strut combination, \strut none}
        \end{axis}
        \end{tikzpicture}
        \vspace{-1.0em}
        \caption{In case of two or more measures, which combination is preferred? Sorted by number of participants preferring the respective combination.}\label{fig:preferred-combination}
        \vspace{3em}
      \end{figure}
      
      In Question 1 of Group 3, we included ``A combination of two or more''. To further understand the most desirable combinations, we asked the participants to select the combination of their choice. 
      %The responses to this question suggest that the most desirable combination is (PIN, Voice) followed by (PIN, Notification).
      In case two or more measures are to be combined, the largest group of participants preferred the combination of ``PIN'' and ``Voice authentication'' (c.f., \Cref{fig:preferred-combination}). Out of the five most popular combinations, four included ``Voice authentication'', signifying that ``Voice authentication'' might be a valuable and non-disruptive additional security and personalization measure.

      \subsection{Security and Personalization Measure Frequency - Group 3, Question 3, and 4}
      
      \begin{figure}[t]
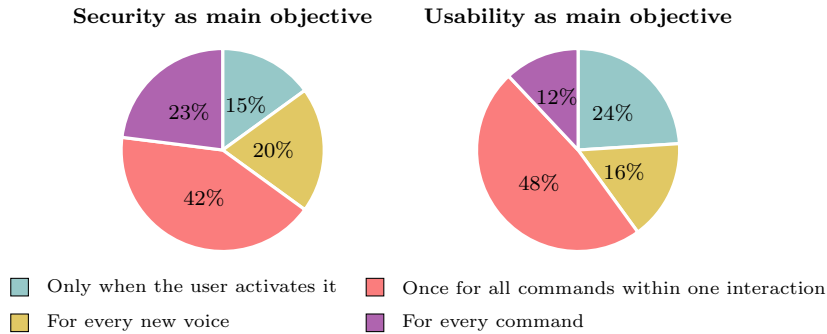

        \centering
\makedoublepiefrequency{Security as main objective}{15/activate,20/newvoice,42/oncecommand,23/everycommand}{Usability as main objective}{24/activate,16/newvoice,48/oncecommand,12/everycommand}
%     slice type={activate}{slightlycolor},
%    slice type={newvoice}{moderatelycolor},
%   slice type={oncecommand}{verycolor},
%    slice type={everycommand}{extremelycolor},
        \caption{How often should the measure be applied, if security/utility was the main focus, respectively?}\label{fig:frequency}
      \end{figure}
      
      In these two questions, we asked the participants to respond how frequently the app should ask for or check the security and personalization measure if the main objective is  security or personalization. \Cref{fig:frequency} summarizes the responses to these two questions. It seems that in both cases, most participants found it sufficient to activate the security and personalization measure once for all commands belonging to one interaction (e.g., speaking the PIN once for all commands within one interaction with a banking app). The participants believed the security and personalization measure should be activated more frequently when security is the main objective compared to when personalization is the main objective.  
      
      \subsection{Open-ended Questions- Group 4}
      
      This group of questions was mainly used as a sanity check to make sure the participants responded to the study with a clear understanding of security and personalization requirements. In response to naming and describing an app for which security could be the main objective, they mentioned apps such as \textit{buying things online, banking and anything financial,  activating and changing device setting, and medical relevant apps}. As an example of apps that would benefit from personalization, the participants mentioned \textit{kids radio and shows, personalizing drink recipes, calendar, music, and adjusting temperature based on one's preference}. In our analysis, we included those participants who showed understanding of the study and discarded the participants who neglectfully answered these questions. 
      This sanity check helped us to ensure that the results presented in this paper are based on the 100 participants who were familiar with smart speaker devices and paid attention to the study.
      
       \begin{comment}

            - Security Concerns
            - Security Requirements (for each Category)
            - Importance of Security for each Category
            - security and personalization Measure Options (Security being the main priority)
        \subsection{Customization Feedback}
            - Customization Expectations
            - Customization Requirement (for each Category)
            - Importance of Customization (for each Category)
            - security and personalization Measure Options (Customization being the main priority)
  
        \subsection{Context-specific}
            how does the result differs from users and non-users, primary and secondary user, and people from different household settings
        \subsubsection{Security and Customization Feedback}
\end{comment}

    %\subsection{Limitation}
    %\subsection{Lessons Learned}

\subsection{Apps with Financial or Payment Implication}
\begin{figure}[t]
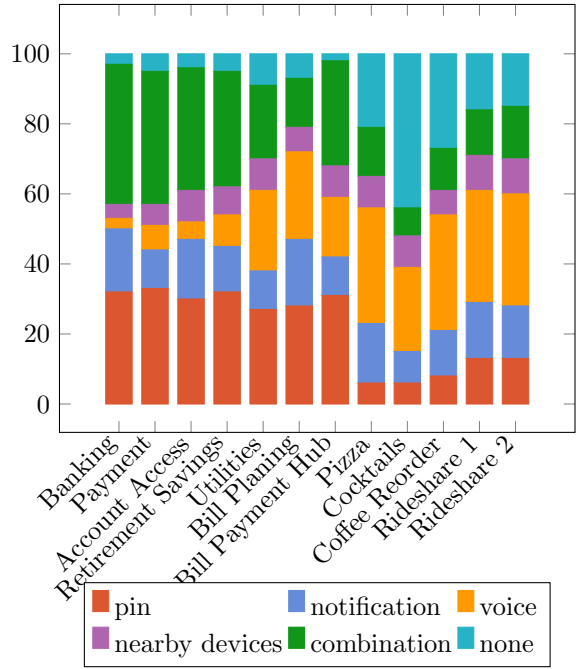

\centering
\makestackedbarchart{Banking, Payment, Account Access, Retirement Savings, Utilities, Bill Planing, Bill Payment Hub, Pizza, Cocktails, Coffee Reorder, Rideshare 1, Rideshare 2}
  {(Banking,32) (Payment,33) (Account Access,30) (Retirement Savings,32) (Utilities,27) (Bill Planing,28) (Bill Payment Hub,31) (Pizza,06) (Cocktails,06) (Coffee Reorder,08) (Rideshare 1, 13) (Rideshare 2, 13)}
  {(Banking,18) (Payment,11) (Account Access,17) (Retirement Savings,13) (Utilities,11) (Bill Planing,19) (Bill Payment Hub,11)  (Pizza,17) (Cocktails,09) (Coffee Reorder,13) (Rideshare 1, 16) (Rideshare 2, 15)}
  {(Banking,03)  (Payment,07) (Account Access,05) (Retirement Savings,09) (Utilities,23) (Bill Planing,25) (Bill Payment Hub,17)  (Pizza,33) (Cocktails,24) (Coffee Reorder,33) (Rideshare 1, 32) (Rideshare 2, 32)}
  {(Banking,04)  (Payment,06) (Account Access,09) (Retirement Savings,08) (Utilities,09) (Bill Planing,07) (Bill Payment Hub,09)  (Pizza,09) (Cocktails,09) (Coffee Reorder,07) (Rideshare 1, 10) (Rideshare 2, 10)}
  {(Banking,40) (Payment,38) (Account Access,35) (Retirement Savings,33) (Utilities,21) (Bill Planing,14) (Bill Payment Hub,30)  (Pizza,14) (Cocktails,08) (Coffee Reorder,12) (Rideshare 1, 13) (Rideshare 2, 15)}
  {(Banking,03)  (Payment,05) (Account Access,04) (Retirement Savings,05) (Utilities,09) (Bill Planing,07) (Bill Payment Hub,02)  (Pizza,21) (Cocktails,44) (Coffee Reorder,27) (Rideshare 1, 16) (Rideshare 2, 15)}
\caption{Preferred security measure per app over all participants.}\label{fig:threecategories}
\end{figure}

We selected 12 apps that directly allow the users to make payments, purchase items, order services, or access financial information. Access to such application might have financial consequences and, if not authorized by the user, could create unwanted financial loss. For example, an adversary might access the smart speaker by being in proximity of the device and transfer the remaining balance of the user to their account. Such an adversary might order a ride or order an item to be delivered to their desired address and charge the user for such transactions. \Cref{fig:threecategories} shows the participants preferred security and personalization measure for the selected applications. 

We see that for the majority of such applications in the Finance and Bill categories, ``PIN'' and ``A combination of two or more'' were the preferred measures. Interestingly, for the apps with shopping and service ordering functionality, while the majority of participants preferred to have \emph{some} measure enabled, ``PIN'' was much less popular than ``Voice authentication''. Here, a cocktail app was an outlier in our analysis, showing that 44\% of the participants selected ``None'' as the preferred security and personalization measure. We suspect that these participants focused on the recipe aspect of the cocktail app, not realizing that this app can also directly make purchases.
%This result shows that

Out of these 12 applications, 2 of the applications (i.e., Payment and Utilities) require the users to set a 4 digit PIN to run certain commands and 2 of the applications (i.,e, Banking, Retirement Savings) let the users select an optional 4 digit PIN.   
7 other applications (i.e., Account Access, Bill Payment, Bill Payment Hub, Pizza, Cocktail, Coffee Reorder, and Rideshare 1), do no have any security and personalization measure in place; anyone accessing the smart speaker can thus learn account information, order items, or request services through them. Finally, Rideshare 2 offers an option to switch between accounts to create a more personalized experience for different household members who share the same smart speaker device.

\subsection{A Closer Look at Offered Measures}
An important question that would drive future studies is to understand whether the selected apps (particularly those in highly relevant categories) meet the participants' expectations for security and personalization measures (e.g., an optional or mandatory PIN). 
To get a picture of the existing measures we manually reviewed the technical description of the apps we considered in our study. %on the smart speaker app store and the app providers' website. 
For apps in the highly relevant categories, we installed each app on a smart speaker and created and linked an app account where account linking was required by the app provider. 
Account linking is required for most of the applications. However, once the account is linked and the app is enabled, anyone accessing the smart speaker can launch the app and execute commands.

The only deployed measure we encountered is a 4 digit PIN.  Out of the 51 apps in the study, only 8 apps ask the users to set a 4 digit PIN and 2 of the apps mention optional PIN setup. 1 app is in the Finance category, 1 app is in the Utilities/Bills category, 1 app is in the Smart Devices category; all 5 apps in the Connected Cars category also had this feature. The results as shown in \Cref{fig:measures-overall}, suggests that for all apps in the Finance category the participants preferred security measures, with the largest fraction preferring ``PIN'' or ``A combination of two or more''; thus, only 1 of the apps in the Finance category somewhat meets the participants' expectation of the type of security and personalization measures. For the rest of the 10 apps that have deployed optional or mandatory PIN, the primary desired  measure in fact is not PIN. For example, in the Connected Cars category the primary method is ``Voice authentication''' followed by ``A combination of two or more''. ``PIN'', ``Notification'' and ``Nearby devices'' are almost equally desirable after ``Voice authentication'' and ``A combination of two or more''. 

Only one app that is in the Travel category allows users to define household members and switch between the household members for a personalized experience. Nevertheless, anyone can switch between household members and no security measure is in place. 

None of the apps in the less relevant categories have deployed  measure, which somewhat meets the majority of the participants' expectation. 

While most apps have not deployed any security and personalization measures, one reasonable ask is for the apps to inform the consumers of the risks associated with enabling the apps or the PIN being exposed. Surprisingly only a few apps have included informative information in their app description. For example, a couple of apps in Finance and Bills category give a short warning to the consumers that while using the app their account information may be read aloud.  
%% add something that the PIN is read aloud and can be learned, the user may forget the PIN, smart speaker device stored the PIN as part of the speech unencrypted. 
Our results suggest that apps (and particularly those in the highly relevant categories) should revisit their descriptive information and the choices of measures in response to users' security and personalization expectations.

\TODOS{Write about the categories of Finance, Bills, and Shopping, zooming in on our findings for those.}

\TODOS{Copy Figure 4 (Preferred security and personalization measures) but selected for apps in these categories. We might either have one pie chart, or we might have one per category}

\TODOS{Fix figure 5} \TODOS{Make Tikz Bar Graphs for the remaining plots instead of these beautiful excel plots}
\TODOM{add the payment context to abstract and intro}
\TODOM{address other todos and remove comments and pointers}

\ifdraft
\begin{figure*}[h]
\centering
\includegraphics[scale=0.5]{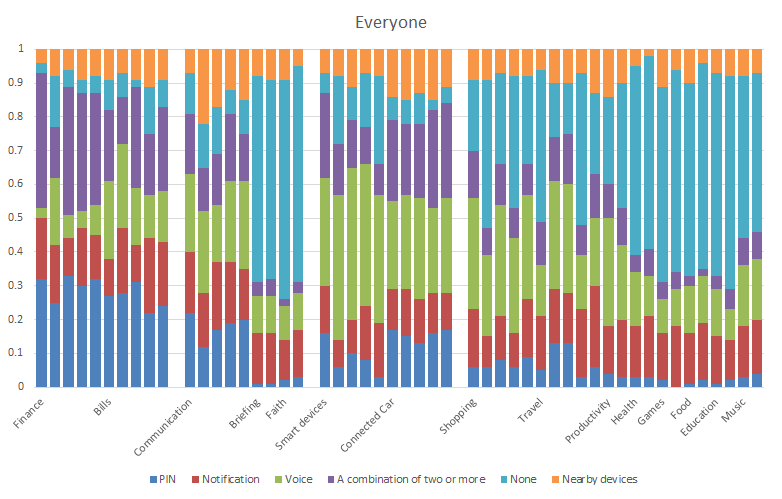}
\caption{Preferred security measure per app over all participants.}\label{fig:options}
\end{figure*}
\fi

\ifdraft
\begin{figure}[h]
\centering
\makesinglepiemeasures{Top 5 Apps}{29/pin,14/notification,12/voice,33/combination,9/nearby,3/none}
\caption{Top 5.}\label{fig:preferred-top5}
\end{figure}
\fi

\ifdraft
\section{Discussion and Countermeasures}
- Voice authentication
- PIN and passphrases
- Second factor authentication 
- continuous authentication
- Extension to other devices

Some of the countermeasures we explored in the study, such as notification and nearby devices, are not widely available today. The responses indicate that especially for security-critical applications the participants did not appear to trust nearby devices as a valid measure: only 9\% of participants selected this measure for apps in categories for which they considered security important or very important, c.f., \Cref{fig:measures-skills-security-important-vs-personalization-important}, which was lower than the number of participants selecting ``None''.

\ifdraft
One cannot but wonder: Does the selected app fulfill the participants expectations 
\fi

\TODOM{Potential mitigation}

\fi

\ifdraft
\subsection{Limitations of this study}

We only considered 50 apps: for each category only 2 or 5 apps were considered to avoid study fatigue. Future studies could focus on individual categories to explore them in greater detail.

Family composition: future studies could focus on how exactly the living situation factors into this.

We only studied one device. However, other devices offer similar third-party applications and offer similar security- and personalization features.

While reviewing apps we noticed that some of them appeared to be outright malicious. This information, however, was sometimes not obvious from the respective app's description but only noticeable from the reviews given for that respective app. If we pair this with the lack of good reviews for many apps (even popular ones), both this and future studies face a challenge in vetting apps to include.

\TODOS{Do we want to mention: We did not consider remote attackers for our study. The only usage we considered was people with direct physical access to a device.}

\fi

\section{Conclusion}
In this paper, we present an extensive study with 100 participants on the importance of security and personalization of third-party apps for smart speaker devices. Our study shows that the majority of the participants had a reasonable understanding of the app categories that demand security or personalization and that they expect to see at least one form or a combination of security and personalization measures (e.g., ``PIN'', ``Voice authentication'') to secure or personalize apps in highly relevant categories. According to our study only 6 apps under the study require the consumers to set 4 digit PIN and hence the participants expectation is largely disregarded in highly relevant categories. We believe the results of our study bring attention to the community to devise secure and usable security and personalization measures to fill the gap between the consumers' expectation and the real world deployment of the apps. 

\TODOS{Summary of results}

%\clearpage
\bibliographystyle{plain}
\bibliography{skills}

\ifdraft
\appendix

\section{List of app categories and descriptions}\label{sec:list-of-categories}
\begin{itemize}
    \item Finance: access your financial information, receive and transfer funds, get balance, access your transactions, stock prices, budgeting.
    \item Utilities/Bill: pay bills, balance info, schedule payment, outage info
    \item Communication: access your contacts, call and send messages, read notifications, emails.
    \item Briefing: listen to news and briefings, weather forecast
    \item Faith: Quotes of the day, listen to inspirational speech, scripture, verses 
    \item Smart devices: access your connected devices, set temprature, lock/unlock door, turn on/off camera, read sensors
    \item Connected Cars: start/stop car engine, lock/unlock car door, get odometer/fuel/temp/location information
    \item Shopping: order items, track orders, sales and coupons
    \item Travel: get schedules, order ride, access ride history, get direction, route info 
    \item Productivity: calendar, lists, reminders, count downs
    \item Health: workout routines, meditations, calming songs, sleep aid, general health and medical knowledge
    \item Games: trivia, games, fun facts, Quizzes, jokes and humour, sport teams fans
    \item Food: food recipes, cocktail recipes, cooking temperature, cooking time, diet, nutrition info
    \item Educational: facts, learning a skill, getting advise, personal growth, common knowledge
    \item Music: playing musics, podcasts and audio books, relaxing sounds
\end{itemize}
\fi

\ifappnames
\section{List of apps}\label{sec:list-of-skills}
\subsection{Most relevant categories}
\paragraph{Finance}
\begin{itemize}
    \item Capital One: Ask Capital One about your credit card, checking, savings, and auto loan accounts. Then try out our new feature, "How much did I spend?" to get quick answers on how much and where you are spending.
    \item YNAB: This is a companion app for YNAB. It requires that you first set things up with YNAB at YouNeedABudget.com. Once you have have a YNAB account, have enabled the app and linked your YNAB account, you are ready to budget like a pro.
    \item Paypal: Want to send or request money and get updates within the United States using only your voice? With the PayPal app, we’re making it easier than ever to check your account updates, including you PayPal balance, and send or request money from friends using nothing more than your voice.
    \item ICMA-RC Account Access: The ICMA-RC Account Access App will allow you to link your Amazon device to your ICMA-RC Retirement Account to quickly hear your account information.
    \item Prudential Retirement: Stay connected with your retirement savings accounts with the Prudential app. Get your retirement savings plan account balance, annual rate of return and outstanding loan balance – all with the power of your voice.
\end{itemize}

\paragraph{Bills}
\begin{itemize}
    \item PSE\&G: This is the official app of PSE\&G (Public Service Electric & Gas) in New Jersey. In this app, customers can link their account then access their billing dashboard through your smart speaker. Hear your balance, due date, and your energy usage details, pay off your balance, schedule service appointments
    \item Bill Planner: Easily keep up with your utility bills with your smart speaker. Get an alert when you’re behind your normal payment date, your bill is higher than normal or your payment has problems clearing. Your smart speaker answers questions about your utility bills and lets you securely pay your utility bill with just your voice.
    \item CoServ: "Here are some voice commands that you can use: 
Account Balance:  What is my balance? What’s my payment amount?
Billing and Payment History : What was my last payment amount? What was my prior billing?
You can make a one-time payment with a previously stored payment option. 
"
    \item DEWA: DEWA customers can interact with DEWA App – AI personal assistant - to enquire about any topics related to DEWA such as: bill enquiry, conservation tips for electricity and water consumptions, details of DEWA happiness centers, payment channels, and much more..
    \item CommEMC: You can request your current balance, get information about your last payment, check for outages on your accounts, read your account number, and get information or alerts. We are always adding new features to better serve you!
\end{itemize}

\paragraph{Communication}
\begin{itemize}
    \item Ask My Buddy: You can ask for alerts, inquire about your current account balance, read your linked accounts, check outage statuses, and read utility alerts.
    \item Cloud Connect: Cloud Connect will call any US or Canadian phone number, announce you to that number, and then connect you to the call.
    \item Phone Link: Phone Link connects your smart speaker to your Android phone or tablet, giving you the ability to read out and send text messages, make phone calls, locate a lost device, and play music and audio files.
    \item Ooma: Check your Ooma Voicemail and control your Ooma Home Security system hands-free. 
    \item Find My Phone: With Find My Phone, you can easily call your lost phone from your smart speaker, without syncing all of your contacts.
\end{itemize}

\paragraph{Smart Devices}
\begin{itemize}
    \item Protection 1: disarm your security system using your voice while your hands are full of groceries walking into the house? To be able to turn on the lights without having to search for the switch? To unlock the front door for a guest while you are in the kitchen? Simply say "ask Protection One to arm my system" and your smart speaker will follow your command.
    \item FireBoard: If you are a FireBoard Drive user, ask your smart speaker to "set my smoker to 225 degrees" or "turn my fan off". 
    \item Home Assistant: "This app is meant for Home Assistant users. When logged into the Home Assistant cloud, you can control all your Home Assistant devices from any compatible device.
Turn your lights on, turn your TV off and enable automations."
    \item Razer Synapse: Razer Synapse allows you to control all your Chroma enabled devices and more using the power of voice. Synchronize lighting, launch programs, perform specific tasks and more across all your devices, smart lighting, and 3rd party RGB devices.
    \item Dyson: You can automate multiple actions using single voice commands of your choice. These are known as Routines and can be things like switching your robot vacuum to quiet mode while you watch the television, just by saying “TV on”. Or setting your air treatment machine to ‘Night Mode’ together with turning off lights, simply by saying “goodnight”.
\end{itemize}

\paragraph{Connected Car}
\begin{itemize}
    \item Lexus: From the convenience of your smartphone or the comfort of your home, you can ask your smart speaker to start or stop the engine, lock/unlock the doors, check fuel level, and check your vehicle’s odometer. You can also manage multiple vehicles in your household by changing your active vehicle and sending commands to it. 
    \item Fordpass: FordPass users with SYNC Connect compatible vehicles can use the Voice Service to issue a variety of remote commands and obtain vehicle information, such as vehicle range and odometer reading. 
    \item Lincoln Way: Keep connected to your Lincoln Connect vehicle with the Lincoln Way app. Lincoln Way users with Lincoln Connect compatible vehicles can use the Smart Speaker's Voice Service to issue a variety of remote commands and obtain vehicle information. 
    \item myChevrolet: "The myChevrolet app allows you to:
Remotely start and stop your vehicle’s engine, 
Remotely lock and unlock your vehicle’s doors,
Manage your household’s other eligible Chevrolet, Cadillac, Buick and GMC vehicles"
    \item myGMC: "The myGMC app allows you to:
Remotely start and stop your vehicle’s engine,
Remotely lock and unlock your vehicle’s doors,
Manage your household’s other eligible GMC, Cadillac, Buick and Chevrolet vehicles
"
\end{itemize}

\paragraph{Shopping}
\begin{itemize}
    \item Domino's: your smart speaker is making ordering your Domino’s favorites even easier. With the Domino’s app, you can build a new order from scratch, place your Easy Order or your most recent order. You can ask your smart speaker for your order’s status with Domino’s Tracker®. 
    \item Sparkling Ice Maker: Get step-by-step instructions on how to make adult-friendly cocktails created by Sparkling Ice's mixologist. This app makes purchases. Your smart speaker allows purchasing by voice using your default payment and shipping settings.
    \item Starbucks Reorder: "Use the Starbucks Reorder app to:
Reorder your Usual from one of the last 10 stores you’ve ordered from, 
Check your primary Starbucks Card balance, 
Switch between your last 5 previous orders"
    \item Send Me A Sample: Whenever you see a brand featuring the Send Me a Sample logo, just say “ask Send Me a Sample” and your product sample of choice will be delivered straight to your door, completely free
    \item Frontgate: Use your smart speaker to ask questions about all your Frontgate orders. Get status updates anytime you like with the voice-activated convenience you love from your smart speaker.
\end{itemize}

\paragraph{Travel}
\begin{itemize}
    \item Washington State Ferry: Don't miss the boat! Get the next departure times for any route in the Washington State Ferry system. If you can't make it, ask for the ferry after that. You can also get information about remaining spaces, and the last scheduled ferry for the day.
    \item Lyft: Need a lift? With the Lyft app, you can request a ride and get picked up in minutes from the home address set in your Lyft account. Just request a car with your voice.
    \item Uber: Get a reliable ride in minutes with the Uber app. Simply say “ask Uber for a ride” and an UberX will be there in minutes. If you would like a different car type, you can say things like "ask Uber to order an Uber Black." From low-cost to premium, every ride option feels like an upgrade to the everyday.
    \item NYC Subway: This is an app that checks the status of NYC's subway lines. You can get a full update of all lines, which tells you which lines are not currently 'Good Service' or check for a status of a specific line.
    \item Flight Service: Leidos Flight Service provides METARs, TAFs, AFDs for airports requested by ICAO codes. Link your 1800WxBrief.com account to receive an updated brief on your filed flight plan before you depart.
\end{itemize}

\subsection{Less relevant categories}

\paragraph{Briefing}
\begin{itemize}
    \item The Daily Show: Trevor Noah and The World’s Fakest News Team break down the day's most relevant headlines. Ask The Daily Show for show highlights, interviews, upcoming guests, and tune-in times. 
    \item The Weather Channel: The Weather Channel app provides accurate forecasts from the world’s most trusted weather source, with information about temperature, chance of rain and more – along with severe weather alerts and radar (for smart speaker devices with screens).
\end{itemize}

\paragraph{Faith}
\begin{itemize}
    \item Examining The Scriptures Daily: "My Daily Text in your smart speaker is for everyone who wants to start a day the right way.
Many find it advantageous to listen to it in the morning. Then you can reflect on such thoughts throughout the day. This app uses content from the Jehovah's Witnesses website."
    \item Catholic Daily: Whether you didn't make it to Mass this week, or you just want time for some spiritual reflection, Catholic Daily provides a piece of the Mass to you from the convenience of your enabled device. Hear all of a day's readings or just the gospel direct from the Roman Catholic Lectionary for Mass. 
\end{itemize}

\paragraph{Productivity}
\begin{itemize}
    \item OurGroceries: "This app lets you add items to the shopping lists on your OurGroceries app. The app keeps your grocery lists instantly synchronized on all the iPhone and Android smartphones in your household.
"
    \item AnyList: AnyList is the best way to create and share grocery shopping lists, and this app allows you to add items to AnyList. After linking your AnyList account, the Shopping & To-do lists will appear as new lists in AnyList.
\end{itemize}

\paragraph{Health}
\begin{itemize}
    \item Guided Meditation: A meditation app by Stop, Breathe & Think featuring over 70 guided meditations designed to reduce the stress and anxiety of daily life. Each day, the "Guided Meditations" app offers a new meditation of the day that is between 1 and 9 minutes long. Over 70 different guided meditation, mindfulness and breathing exercises are included so each day you get a fresh meditation.
    \item Five Minute Workout: Five Minute Workout: Core and Cardio is a fun fast moving 5 minute routine packed with fat burning exercises that will tone your core right into shape. It's the most bang for your buck that you can get when it comes to working out.
\end{itemize}

\paragraph{Games}
\begin{itemize}
    \item Fetch Facts: On Wednesdays we play trivia! Test your MEAN GIRLS knowledge with this voice app. Featuring nearly 50 fetch facts about both Broadway musical the 2004 feature film from Tina Fey.
    \item NHL: The Official NHL App is the best way for hockey fans using the smart speaker to get detailed information on NHL scores, schedules, standings and current players.
\end{itemize}

\paragraph{Food}
\begin{itemize}
    \item Happy Hour: This app will give recipes for over 170 popular bar drinks.
    \item Whole30: The Whole30 is a 30-day diet that emphasizes consuming whole foods and eliminating sugars, alcohol, grains, legumes, soy, and dairy from your diet.
\end{itemize}

\paragraph{Education}
\begin{itemize}
    \item Magoosh Vocabulary Builder: All the words in this app have been handpicked by experts who prepare students for the GRE, GMAT, SAT and ACT. You will start at the apprentice level and progress to becoming a vocabulary master.
    \item Kids Quiz: Kids Quiz is a great way for children to reinforce their learning and discover new things about the world. 
\end{itemize}

\paragraph{Music}
\begin{itemize}
    \item SiriusXM: Listen to 300+ channels of great SiriusXM® programming -- live and on demand -- including 100+ Xtra music channels and create Personalized Stations Powered by Pandora! 
    \item Pandora: Explore the new Pandora. From the free stations you love to on-demand songs and albums, Pandora gives you a personalized experience tuned to the moment you’re in.
\end{itemize}

\subsection{Selected Application with Financial or Payment Outcome}
\label{sec:finpay}
\begin{itemize}
\item Capital One
\item Paypal
\item Account Access
\item Prudential Retirement
\item PSE\&G (Public Service Electric & Gas)
\item Bill Planner
\item CoServ
\item Domino's
\item Sparkling Ice Maker
\item Starbucks Reorder
\item Lyft
\item Uber
\end{itemize}

%%%%%%%%%%%%%%%%%%%%%%%%%%%%%%%%%%%%%%%%
\else % NO APP NAMES
\fi

\end{document}